\documentclass[prb,amsmath,amssymb,reprint]{revtex4-2}

\usepackage{amssymb}
\usepackage{amsmath}
\usepackage{dcolumn}
\usepackage{bm}
\usepackage{color}
\usepackage{graphicx}

\usepackage[justification=raggedright]{caption}

\usepackage{subcaption}

\begin{document}

\title{
Superconducting spin valves based on a single spiral magnetic layer
}

\author{N. G. Pugach}

\affiliation{HSE University,
101000, Moscow, Russia}

\author{M. O. Safonchik}
\affiliation{A. F. Ioffe Physical-Technical Institute,  RU-194021, St
Petersburg, Russia}

\author{V. I. Belotelov}

\affiliation{Lomonosov Moscow State
University, Leninskie Gory 1(2), 119991, Moscow, Russia}

\affiliation{\mbox{V.I. Vernadsky Crimean Federal University, Vernadsky Prospekt, 4, Simferopol 295007, Crimea}}

\author{T. Ziman}
\affiliation{Institut Laue-Langevin, BP 156, 41
avenue des Martyrs, 38042 Grenoble Cedex 9, France}

\author{T. Champel}
\affiliation{Univ. Grenoble Alpes, CNRS, LPMMC, 38000 Grenoble, France
}

\begin{abstract}
A detailed investigation of a superconducting spin-triplet valve is presented. This spin-valve consists of a superconducting film covering a metal with an intrinsic spiral magnetic order, which could result  from  competing isotropic exchanges or, if the crystal lattice breaks  central symmetry, from  asymmetric Dzyaloshinskii-Moriya exchange.
Depending on the anisotropy, such a metal may change its magnetization  either from a spiral to  uniform order, as seen in Ho and Er,
 or in the direction of the spiral itself, as in crystals of the B20-type structure (such as  MnSi, (Fe,Co)Si, FeGe, etc.).
The nonuniform
magnetic order controls the appearance of long-range triplet superconducting
correlations at strong exchange fields,
affecting  the detailed character of the
proximity effect.
We show that the magnetic control of the spin-valve behavior can also be obtained from moderately low exchange fields (typically associated to negligible long-range triplet correlations),
thanks to an orientation-dependent averaging mechanism of the magnetic inhomogeneity on the scale of the Cooper pairs.
Our numerical calculations reveal that the spin-valve effect is in fact magnified at moderately low exchange fields, when the exchange splitting in the spiral magnet is comparable to the superconducting gap, and the spiral period is less than or equal to the superconducting coherence length in the magnet multiplied by $2\pi$.

\end{abstract}

\date  \today

\maketitle

\section{Introduction}

Superconducting spintronics is a modern field in cryogenic nanoelectronics \cite{Roadmap2018} that has emerged since  the second decade of this century \cite{Eschrig2011}. As in traditional spintronics, the aim is to utilize spin transfer for
information processing.
Since a spin current is not necessarily  accompanied by a charge transfer, such devices, combining magnetic and superconducting order, promise to be energetically efficient. This field is currently in development,  with the demonstration of device concepts \cite{Robinson2014,Linder2015,Eschrig15b}. Magnetic control of charge transfer is a key mechanism of spintronic devices and, in particular, of the functionality of magnetic memory (MRAM) units.  Like traditional MRAM elements based on magnetic tunnel junctions and spin transfer torque elements, superconducting spin valves (SSVs) were  first proposed theoretically \cite%
{Beasley1997,BuzdinVed1999,Tagirov1999}, and are currently
at the stage of optimization \cite%
{Aarts2015,Garifullin2016,Birge2018a}.

The magnetization reversal of a ferromagnetic (F) element  may provide for magnetic control of the Josephson current \cite{Vedyayev2005,Kupriyanov2002,Ryazanov2012,Birge2018b} or may change the superconducting critical temperature $T_c$ both in the F/S/F \cite{Fominov2003} and S/F/F \cite%
{Fominov2010} configurations, where S denotes a superconducting film. The possibility of tuning superconductivity by the mutual orientation (parallel or antiparallel) of two F layers in this type of SSV was shown in different experiments
\cite{Gu2002,Potenza,Westerholt2005,Steiner}.
The shift $\delta T_{c}$ in the critical temperature was assumed to be due to the paramagnetic effect, {\it i.e.}, the breaking of the Cooper pairs by the spin polarizations induced within the superconducting layer by the ferromagnetic  layers. When the magnetizations in the F layers are antiparallel, these polarizations effectively cancel out, while in the parallel configuration the induced polarizations act cooperatively to suppress superconductivity.
A shift in $T_c$  between two different magnetic configurations opens the way to  realize an optimized switch from a superconducting to a normal state. The formal value of the magnetoresistance is then infinite, since the resistance changes from  zero in the superconducting state, to a finite value in the normal state.

The triplet mechanism of the SSV effect was later deduced  in the same structures. It was shown that a noncollinear magnetization configuration may provide an even larger shift in  $\delta T_{c}$
\cite{Flokstra2015,Aarts2015,Garifullin2013,Blamire2014} due to the appearance of long range triplet superconducting correlations (LRTC) \cite%
{BVE:PRB2001,Shekhter,Eschrig03,VolkovRevModPhys}. The production of the LRTC, which can be seen as  one more channel for the draining of the Cooper pairs from the superconductor, enhances the proximity effect, and as a result reduces $T_{c}$ more efficiently. A full $T_{c}$ switch, where $\delta T_{c}$ is larger than the superconducting transition width, has been reached in  a few experiments \cite{Leksin2010,Leksin2012,Li2013,Gu2015}. Nevertheless, the improvement and
optimization of SSVs still remain both a theoretical and experimental challenge
\cite{Garifullin2018,Brazil2017,Lenk2017,Baker2018,Valls2018,Robinson2018,Kupriyanov2018}%
. Possible ways to improve the properties of SSVs include optimization of the structures, via the technology of their  preparation, the quality of interfaces, and the choice of materials, including strong, half-metallic, or insulating ferromagnets.

Recently, layers of the spiral magnet (M) holmium have been used as spin-mixers, improving the magneto-resistive properties of SSVs \cite{Sosnin2006,Zhu2013,Blamire2014b,Blamire2015,Aarts2015,Chiodi2013,DiBernardo2015}%
. Taken separately in a S/M bilayer, a spiral or conical magnetic layer also
provides a shift of $T_{c}$  when an external magnetic field is applied \cite{DiBernardo2015,Satchell2017}. Indeed, a parallel magnetization of the spiral magnetic layer suppresses the LRTC, leading to a change in $T_{c}$. However, the return to the initial helimagnetic state would require heating the sample above the Curie temperature,  much higher than $T_{c}$. This would make this kind of SSV difficult to use in low-temperature electronics.

A further important problem to consider, inherent both for SSVs
and Josephson MRAM, is the so-called half-select problem \cite{Vernik2013}.
This occurs in the conventionally known RAM scheme with two sets of crossing ``word'' and `''bit'' lines, for example along the ``X'' and ``Y'' directions. The control signal should switch the state of a particular element at the crossing of two lines in
a net, whereas the half-amplitude signal should not cause a switch of any other elements along the crossing lines.
This requirement provides for an addressable switch of any particular memory element in RAM. It means that the two logical states of a memory element must be separated by a potential barrier.
Multilayered structures with continuous magnetization reversals do not solve
this problem. In contrast, it has been pointed out \cite{Pugach2017} that if a spiral magnet M has a potential barrier between two spiral directions due to cubic anisotropy, the resulting S/M bilayer  may work
both as an SSV and a Josephson MRAM. Recently, a way of controlling  such SSVs was developed \cite{Gusev2021},  based on magnonic relaxation after a  magnetic field pulse. This method  is attractive as it is non-destructive of the superconducting state. The spiral SSVs allowing such a means of control should have parameters (record-readout speed and stability) comparable with modern MRAM \cite{Gusev2021}.

In this paper we explore this idea  quantitatively \cite{Pugach2017} by considering spiral magnets of the B20 family, such as  MnSi, (Fe,Co)Si, FeGe, in bulk, or etched films, to provide the full $T_{c}$ switch by a change of the spiral direction. This is possible because  the cubic noncentrosymmetric magnetic structure of the B20 crystal family \cite{Ishikawa1977,Pfleiderer2001,Uchida2006}
 provides an asymmetric Dzyaloshinskii-Moriya exchange \cite{Dyadkin2}.
 The magnetic spiral structure, characterized by the spiral vector $\mathbf{Q}$, may be aligned in a few different, but equivalent, directions under the control of a weak external magnetic
field. These preferred directions are determined by the potential barriers that
depend on the compound \cite{Dyadkin}.
This provides the  advantage previously mentioned  for their use as MRAM
switchable elements. Such magnetic compounds and their films are now the subject of much interest \cite%
{Muhlbauer2009,Fert2013} as a medium for magnetic topological defects like
skyrmions. Thin films of MnSi have a strong in-plane magnetic anisotropy and
host only magnetic spiral order with the vector $\mathbf{Q}$ perpendicular to the plane of the film \cite{Menzel}.
The $T_{c}$ in such a superconducting bilayer  may be controlled via  a reversible uniform magnetization of the films in an external magnetic field. Such elements may be used in biased RAM. Here we  investigate theoretically the  possibilities of using S/M bilayers as switchable elements for cryogenic electronics application, useful for both SSVs and MRAM. We find that the magnitude of $\delta T_c$ is a non-monotonic function of the spiral wave vector  that depends on the amplitudes of the exchange field. This behavior can be exploited for the optimization of the structures.

This paper is organized as follows: Sec. \ref{Model} presents the model and the general method for deriving $T_c$ in the S/M bilayer, while Sec. \ref{Discussion} discusses the numerical results for $T_c$ we obtain. Sec. \ref{Conclusion} concludes this work. The details of the analytical calculations are presented in the Appendix.

\begin{figure}[t]
\begin{center}
\includegraphics[width=0.4\textwidth]{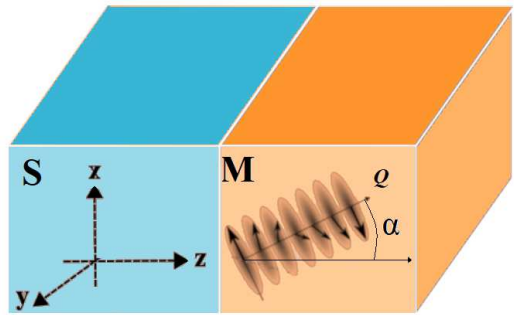}
\end{center}
\caption{A sketch of the studied SSV with the spiral wave vector $%
\mathbf{Q}$ tilted at an angle $\protect\alpha$ to the normal of the S/M interface.}
\label{Sketch}
\end{figure}

\section{Model}

\label{Model}

To describe the superconducting proximity effect with a spiral
magnet, we use linearized Usadel equations in the form \cite%
{Champel2005} valid close to $T_{c}$ in the dirty limit, which is usually satisfied
in hybrid structures prepared by sputtering. In this limit, the
superconducting coherence lengths in the M and S layers are given by $\xi
_{f,s}=\sqrt{D_{f,s}/2\pi T_{cb}}$, where $D_{f,s}$ are the corresponding
electron diffusion coefficients and $T_{cb}$ is the critical temperature of
the bulk superconductor.

Here we have developed a method of  calculation that  allows us to consider the  limits of both weak and strong exchange splitting $h$, as well as to examine the behavior at arbitrary values of $h$ (provided the Usadel equations remain valid) for  spirals of both long and short periods. Moreover, we take the  thickness of the magnetic layer to be finite, and inlcude related effects, such as re-entrant superconductivity, which also depend on the spiral magnetic order. The third improvement of the current method in comparison with previous theories \cite{Champel2005,VolkovAnishchanka,Halasz2009,Halasz2011,Wu2012,USA2014,Halterman2016,Pugach2017} is the determination of $T_c$ at arbitrary angles between the spiral wave vector $\mathbf{Q}$ and the plane of the S layer.

\subsection{Usadel equations for the magnetic spiral inclined at an arbitrary
angle $\alpha$}

We consider a superconducting layer (S) with a finite thickness $d_{s}$ in proximity with a layer of spiral magnet (M) of thickness $d_{f}$. The spiral vector $\mathbf{Q}$, is taken to be inclined with an angle $\alpha$ with respect to the $OZ$ direction, orthogonal to the plane of the layers, see Fig. \ref{Sketch}.  We assume that $\mathbf{Q}$ lies in the $XOZ$ plane, while the $XOY$ plane coincides with the SM interface \cite{note}.

The linearized Usadel equations have the general form \cite{Champel2005,Champel2005b}
\begin{eqnarray}
\left( D\nabla ^{2}-2\left\vert \omega \right\vert \right) f_{s} &=&-2\pi
\Delta +2i\ sgn(\omega ) \, {\bf h}\cdot {\bf f}_{t},
\notag \\
\left( D\nabla ^{2}-2\left\vert \omega \right\vert \right) {\bf
f}_{t} &=&2i\ sgn(\omega ) \, {\bf h} \, f_{s},  \label{UU0}
\end{eqnarray}
where the singlet $f_{s}$ and triplet $\mathbf{f}_{t}=(f_{x},f_{y,}f_{z})$
spin components of the anomalous Green's function describe the different superconducting correlations. The singlet superconducting order parameter $\Delta = \Delta(z)$ is nonzero only in the S layer, while the exchange spin splitting $\mathbf{h}(x,z)$, aligned along the local magnetization, is nonzero only in the M layer.
The spiral vector ${\bf Q}$ enters in Eq. \eqref{UU0} through the exchange spin splitting, which for $\alpha=0$  reads ${\bf h}({\bf r})=h (\cos [Qz], \sin [Q z], 0)^{T}$.
 By taking into account the even symmetry of the singlet components and the odd
symmetry of the triplet ones with respect to the Matsubara frequency $\omega
\equiv \omega _{n}=\pi T(2n+1)$ with $n$ a positive or negative integer, we may consider in the following only positive $\omega$.

As a result of the presence of the S/M interface and of the proximity effect, the superconducting correlations naturally depend on the distance to the interface, i.e., they exhibit a $z$-dependence.
For an inclined spiral vector ${\bf Q}$, the magnetic vector ${\bf h}$ displays an explicit spatial dependence on both variables $x$ and $z$, so that the problem we consider is in general a two-dimensional one. However, as we show in the Appendix, the extra dependence on $x$ due to the rotating spiral when $\alpha \neq 0$ can be gauged away via a suitable transformation on the triplet vector $\mathbf{f}_{t}$, so that the original problem is mapped to the following effective one-dimensional one for the triplet components $f_{-}(z)$ and $f_{+}(z)$ in the magnetic layer (see the Appendix for details):
\begin{eqnarray}
\left( \frac{\partial ^{2}}{\partial z^{2}}-\frac{2\omega }{D_{f}}\right)
f_{s} =i\ \frac{h}{D_{f}}\left[ f_{-}-f_{+}\right] ,  \hspace*{2.8cm} \notag \\
\left[ \frac{\partial ^{2}}{\partial z^{2}} \mp 2iQ\cos \alpha \frac{\partial }{%
\partial z}-Q^{2}-\frac{2\omega }{D_{f}}\right] f_{\pm} = \mp \, 2i\ \frac{h}{D_{f}}%
f_{s}, \hspace*{0.5cm} \label{UU:2}
\end{eqnarray}
with $h$ and $Q$ the amplitudes of the exchange splitting and of the spiral wave vector, respectively.
The form of these equations reproduces the limiting cases of spirals both orthogonal and parallel to the interface $\alpha=0,\pi/2$ that were considered in Refs. \onlinecite{Pugach2017,Champel2005}.
The presence of linear terms in the spiral vector $Q$ reflects the chirality of the spiral, which is related to the broken inversion symmetry of the magnetic lattice that can be found in compounds of the B20 family, for instance. The chirality makes the calculations rather more complicated than for the case where uniform magnetization can be assumed.

\subsection{The characteristic equation}

If the M layer is semi-infinite and occupies the half-space $z>0$, the
solutions of the linear differential Eqs. (\ref{UU:2}) take the simple form  $f_{j}(z)=u_{j}\exp (-k_{i}z)$, where $j={s,+,-}$. We will take the more general situation of a M layer with a finite thickness. This means that we also have to consider spatial solutions with $-k_{i}$, describing the waves reflected from the free M interface.

The amplitude coefficients for these solutions are determined by the boundary
conditions. Here the wave vectors $ \pm k_{i}$ are the eigenvalues of the system (\ref{UU:2}) and we represent in the following the components of these two related eigenvectors by $u_{j},~v_{j}$. Let us also introduce the characteristic momenta $k_{\omega }= \sqrt{2\omega /D_{f}},~k_{h}=\sqrt{h/D_{f}}$.
Eqs. (\ref{UU:2}) lead to the system of algebraic equations
\begin{eqnarray}
\left( k^{2}-k_{\omega }^{2}\right) f_{s}-i\ k_{h}^{2}f_{-}+i\
k_{h}^{2}f_{+} &=&0,  \notag \\
 \pm 2i\ k_{h}^{2}f_{s}+\left( k^{2}-k_{\omega }^{2} \pm 2iQk\cos \alpha
-Q^{2}\right) f_{\pm} &=&0.  \label{kEq}
\end{eqnarray}

The characteristic equation of this system yielding the eigenvalues $k_{i}$ is
\begin{eqnarray}
\left[ (k^{2}-k_{\omega }^{2}-Q^{2})^{2}+4Q^{2}k^{2}\cos ^{2}\alpha \right]
\left( k^{2}-k_{\omega }^{2}\right) \nonumber \\ +4k_{h}^{4}(k^{2}-k_{\omega }^{2}-Q^{2})=0.
\label{CharEq}
\end{eqnarray}
This bicubic equation displays 3 pairs of solutions $\pm k_{i}$ with $\mathrm{Real} \, k_i>0$, which are the eigenvalues of the system (\ref{UU:2}). In the limit $k_{h}^{2}\gg Q^{2},k_{\omega }^{2}$ the 3 eigenvalues are similar to those found in the
orthogonal case: two short-range components $k_{\pm }\approx\left( 1\pm i\right) k_h$, and one long-range component $k_{0}\approx\sqrt{k_{\omega }^{2}+Q^{2}} $.
Assuming a large exchange energy of $h=100$ meV (it is typically of the order of 1eV in ferromagnetic transition metals),
 one may  estimate the
quasi-momenta  as \cite{Pugach2017} $k_{h}=\xi _{h}^{-1}\sim
1 \, $nm$^{-1},~k_{\omega }\sim 1/7 \, $nm$^{-1}=0.14 \,$nm$^{-1},$ that yields the inequalities $%
k_{h}>Q>k_{\omega }$. The approximation of strong exchange energy for the
values of $k_{\pm ,0}$ is thus valid for $h=100$ meV with an accuracy of two decimal places. The approximate eigenvectors $(-1,-1,1),~(1,-1,1)$ and $(0,1,1)$ associated with these eigenvalues have an accuracy of only one decimal place, however.

In the general case we have solved the  characteristic equation exactly  (\ref{CharEq}), with the eigenvectors so found $u_{j}$ and $v_{j}$ written in the Appendix. Note that, in the approximation of large exchange splitting, the eigenvalues and eigenvectors  coincide with those described previously in the two limiting cases of the spiral parallel to the S/M interface \cite{Champel2005,Champel08} ($\alpha=\pi/2$) and of the spiral orthogonal to the S/M interface ($\alpha=0$) for semi-infinite \cite{Pugach2017} and finite \cite{VolkovAnishchanka} M layers.

The solution of the Usadel equations for the finite M layer may finally be found in the general form
\begin{equation}  \label{GenSol}
f_{j}(z) =\sum\limits_{i}A_{i}u_{j}(k_{i})\exp
(-k_{i}z)+B_{i}v_{j}(k_{i})\exp (k_{i}z).
\end{equation}
The coefficients $A_{i}$ and $B_{i}$ are then determined from the boundary conditions,as discussed in the next Section.

\subsection{Boundary conditions}

Kupriyanov-Lukichev \cite{KL} boundary conditions for the singlet and triplet components of the anomalous Green's function read at the free interfaces
\begin{equation}
\frac{\partial }{\partial z}f_{s,t}=0,  \label{KLBCfree}
\end{equation}%
and at the SM interface located at $z=0$
\begin{eqnarray}
\xi_s \frac{\partial }{\partial z}f_{s,t}^{S} &=&\gamma \xi_f \frac{\partial }{\partial z%
}f_{s,t},  \label{KLBC} \\
f_{s,t}^{S} &=&f_{s,t}-\gamma _{b} \xi_f \frac{\partial }{\partial z}f_{s,t}.
\notag
\end{eqnarray}
We have introduced here  the dimensionless interface parameters $\gamma _{b}=R_{b}A\sigma
_{f}/\xi_f$ and $\gamma =(\sigma _{f}/\sigma _{s})(\xi _{s}/\xi_f)$. The quantities $R_{b}$
and $A$ are the resistance and the area of the S/M interface  respectively,
and $\sigma _{f,s}$ is the conductivity of the M or S metal. Inclusion of the  superscript $S$ in the components $f_s^S$ or $f_t^S$ is used to specify that the correlations are evaluated from the superconducting side, while the absence of superscript holds for the magnetic side.

Since the Usadel equations have been rewritten in terms of the transformed triplet components $f_{\pm}(z)$, we need to derive the boundary conditions for these functions instead. As shown in Appendix, we find the boundary conditions for the free interfaces at $z=d_{f}$ and $z=-d_{s}$ for the inclined case corresponding to the linearized Usadel
equations (\ref{UU:2})
\begin{eqnarray}
\frac{\partial }{\partial z}f_{s}(d_{f}) =  0,  \hspace*{0.5cm}
\frac{\partial }{\partial z}f_{\pm }(d_{f})  = \pm iQ\cos \alpha \,
f_{\pm }(d_{f}),   \hspace*{0.5cm} \label{BCfree} \\
\frac{\partial }{\partial z}f_{s,\pm }^{S}(-d_s)  =  0,  \hspace*{5.3cm} \notag
\end{eqnarray}
and the boundary conditions at the S/M interface at $z=0$
\begin{eqnarray}
\xi_s \frac{\partial }{\partial z}f_{s}^{S}(0) &=&\gamma \xi_f\frac{\partial }{\partial
z}f_{s}(0),  \notag \\
\xi_s \frac{\partial }{\partial z}f_{\pm }^{S}(0) &=&\gamma \xi_f \left( \frac{\partial
}{\partial z}\mp iQ\cos \alpha \right) f_{\pm }(0),  \label{BC:SM} \\
f_{s}^{S}(0) &=&f_{s}(0)-\gamma _{b} \xi_f \frac{\partial }{\partial z}f_{s}(0),
\notag \\
f_{\pm }^{S}(0) &=&f_{\pm }(0)-\gamma _{b} \xi_f \left( \frac{\partial }{\partial z}%
\mp iQ\cos \alpha \right) f_{\pm }(0).  \notag
\end{eqnarray}
The sign $\pm $ in front of the contributions depending on $Q$ expresses the chirality of the magnetic spiral: clockwise or anticlockwise. In the presence of a finite thickness of the magnet, this chirality adds a degree of complexity that was absent  in  previous work. We note though that the superconducting critical temperature turns out to be independent of the sign of the chirality, owing to the spin symmetry of the superconducting
wave function in S.

\subsection{Critical temperature calculation}

The singlet component $f_{s}^{S}(z)$ in the superconductor depends on the superconducting gap $\Delta(z)$, which is calculated self-consistently. The required closed boundary value problem for the singlet component $f_{s}^{S}$ includes the Usadel
equation (\ref{UU:2}) in the superconducting layer, and the boundary condition in the form
\begin{equation}
\left. \xi _{s}\frac{\partial }{\partial z}f_{s}^{S}\right\vert
_{z=0}=-\left. Wf_{s}^{S}\right\vert _{z=0}.  \label{BCW}
\end{equation}
The real-valued quantity $W$ contains the entire information about the proximity
effect with the spiral magnet, and may be written as
\begin{equation}
W=\gamma  \left[\frac{\sum\limits_{i}S_{i}u_{s}(k_{i})}{\xi
_{f}\sum \limits_{i}R_{i} k_i u_{s}(k_{i})}+\gamma _{b} \right]^{-1}, \label{W}
\end{equation}
with the coefficients $R_i$ and $S_i$ defined  in the Appendix.

Finally, we compute numerically the critical temperature $T_c$ of the superconducting layer,  using the self-consistent equation
\begin{equation}
\ln \frac{T_{cb}}{T_{c}}=\pi T_{c}\sum_{\omega =-\infty }^{\infty } \left(%
\frac{1}{ \left\vert \omega \right\vert} -\frac{f_{s}^{S}(z)}{\pi \Delta(z)}
\right)  \label{SelfCons}
\end{equation}
and the method of the fundamental solution \cite%
{Fominov2002,Champel2005,Lofwander07}.

\subsection{Model parameters}

For the realization of the presently studied SSV, different types of spiral magnets may be considered, thus providing various possible ranges of values for the model parameters. This variety can naturally be exploited to optimize the switching control and the spin-valve behavior in the S/M bilayer. In this section, we will address in some detail the possibilities currently  available for real materials and the associated variable ranges, in particular for the magnitudes of the exchange splitting $h$ or for the spiral wave vector $Q$.

Metals of the B20 family like MnSi, (Fe,Co)Si, FeGe, etc., are arranged in a crystal
lattice of cubic symmetry with broken inversion symmetry. The latter provides an asymmetric Dzyaloshinskii-Moriya  exchange, leading to  spiral magnetic order. For Lanthanide metals like Ho or Er, the spiral magnetic order stems from the competing magnetic exchanges of nearest neighbor and next-nearest neighbors. Although the microscopic natures differ, at the level of the proximity effect these magnetic materials exhibit similar properties when described within a mean field approximation.

Because of the lattice symmetry in B20 materials, the magnetic spiral characterized by the vector $\mathbf{Q}$ may align in a few different directions in the ground state, in the absence of an external magnetic field.  Depending on the compound, the preferred spiral directions may be along the cubic axes [100], [010], [001] , as for FeCoSi and MnGe, and along the diagonals of the cube $[1\pm 1\pm 1]$, as for MnSi.
Depending on the magnetic history \cite{Dyadkin}, the spiral may even have an arbitrary direction at a weak cubic anisotropy, as for example in FeCoSi.
The direction of $\mathbf{Q}$ may be controlled by a relatively weak external
magnetic field, of the order of 100 Oe for MnSi.
It is important for the applications in SSVs to have the control field lower
than the critical magnetic field of superconducting films, which is $\sim
30$ kOe for the in-plane direction of the Nb superconductor.
Very recently, a method to control a bilayer SSV which does not perturb its superconducting state  has been proposed and optimized \cite{Gusev2021} via numerical experiments on the MnSi/Nb bilayer SSV.
Switching between a few ground state magnetic configurations  with different directions of the magnetic spiral $\mathbf{Q}$, that are separated by potential barriers,  was proposed  by applying a magnetic field pulse several hundred {\it ps} in duration and  several kOe in magnitude.
Such a pulse would not itself destroy the superconducting state of the Nb layer,  but leads to the excitation of magnons in the magnetic layer, that then trigger the reorientation process of the magnetic spiral  \cite{Gusev2021}. The system can be switched  back and forth by magnetic fields of opposite signs along a single  direction of  the plane of the layers. This allows for easy control. The switching time is estimated \cite{Gusev2021} to be several nanoseconds, which coincides with the scales of the spin-transfer torque MRAM recording time, making this method attractive
for energy saving cryogenic electronics.

The absolute value of the spiral vector is defined as $Q=2\pi /\lambda $, where $\lambda $ is the spiral spatial period. The spiral antiferromagnets of the B20 family may have short or long spiral spatial periods (from 3 nm for MnGe to up to 90 nm for Fe(0.5)Co(0.5)Si)  that reflect the relation between the Dzyaloshinskii-Moriya interaction and the exchange energy.
In the spiral magnetic metal MnSi with $\lambda=18$ nm , the preferred spiral
directions $[1 \pm1 \pm1]$ have the angle $\arccos (1/3)=70.5^{\circ }$. If
two of these directions lie in the plane of the film, the other two make the angle $\alpha =\pi/2-\arccos (1/3)=19.5^{\circ }$ with the normal $OZ$ to the
interface. In this case the structure is periodic in the $OX$ direction with a
large period $\lambda /\sin \alpha =54$ nm, that is much larger than other
characteristic lengths such as $\lambda /2\pi $ or the superconducting coherence length $\xi_f$. The period of the spiral in the $OZ$ direction increases only slightly, since $\lambda /\cos \alpha =19$ nm.

There has been some recent evidence that MnSi has a relatively weak exchange splitting
\cite{Janoschek,Bauer}. In the compound MnSi, the amplitude of the exchange field has even been estimated \cite{Bauer} to take the re value $h=11 $ meV, low  in comparison with ferromagnetic metals.
We could take the estimated value of the exchange
as an upper limit for the exchange splitting of the superconducting electrons in the mean field approximation.
The exchange splitting magnitude may, however,  be larger for other compounds of the B20 family. The Curie temperature may be taken as a signature of the exchange energy magnitude, though without direct proportionality. It varies from about 29 K for MnSi to~279 K for FeGe. Thus, the Curie temperature is larger than the $T_{c}$ of the s-wave BCS superconducting metals like Al or Nb usually used in cryogenic nanotechnology.

As a result of the asymmetric exchange interaction, topological magnetic
defects called skyrmions exist in such compounds at high magnetic fields and
at temperatures close to, but below, the Curie temperature \cite{Neubauer,Li}. These defects carry a nontrivial topological charge. Because of possible
applications in spintronics \cite{Moche}, as well as of the fundamental
interest for physical properties dictated by topology, these compounds and their films are now under very intensive theoretical and experimental investigations.  Films of B20 compounds  can be prepared experimentally from the single crystals by etching \cite{Tonomura,Mochizuki,Yu}.
Despite their different geometry, such films have essentially the same magnetic properties as bulk crystals. In contrast,  thin films of MnSi grown
by molecular beam epitaxy (MBE) or by electron beam lithography often do not appear to host skyrmions, presumably due to their high in-plane magnetic anisotropy. In these structures, the magnetic spiral may align only
with $\mathbf{Q}$ orthogonal to the film plane. At $d_{f}<\lambda $ the incomplete
period of the magnetic spiral may be continuously collapsed towards a uniform
magnetization, replacing the conical phase in a parallel magnetic field from about
few kOe up to around 1.3 T for MnSi. This effect may be also used to tune $T_{c}$ in the S/M bilayer, by gradually destroying the LRTC when approaching the situation of a uniform
magnetization.

In contrast, the spiral magnetic orders in lanthanide 4f metals, such as Ho or Er which have been widely used in experiments on the superconducting
proximity effect,  are characterized by short spirals, of period about 4 nm for Er and 6 nm for Ho. The order of magnitude of the exchange splitting in these materials is unsettled \cite{remark}.
%and probably by a large exchange splitting.
Furthermore, the magnetic anisotropy of such lanthanide films is usually strongly in-plane.

{\em Other model parameters.}
For the $T_c$ calculations, the superconducting layer is assumed to be made of Nb, with the superconducting coherence length $\xi_s=11$ nm. The  critical temperature of the bulk superconductor is taken as $T_{cb}=9.2$ K. The superconducting coherence length in the M layer is chosen to be $\xi_f=4.2$ nm. The S/M interface
transparency is taken from the realistic estimations made in Ref. \onlinecite{PugachPRB2011} with $\gamma_B =0.7$, and the interface parameter \cite{Pugach2017} is chosen as $\gamma=0.7$.

\section{Results and Discussion}

\label{Discussion}

\begin{figure*}[tb]
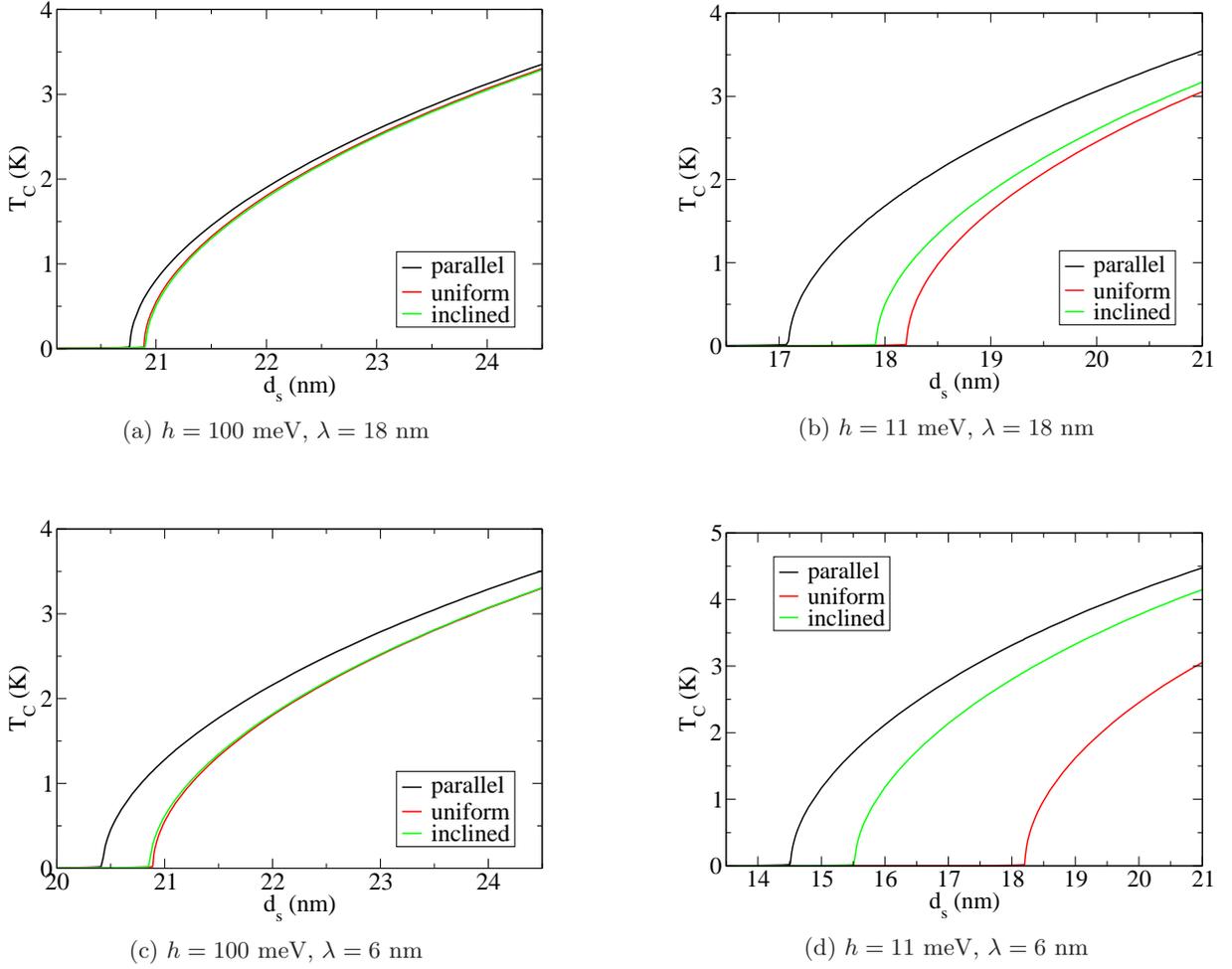

\begin{subfigure}{.5\textwidth}
  \includegraphics[width=.8\linewidth]{Fig2a.eps}
  \caption{$h=100$ meV, $\lambda=18$ nm}
  \label{Fig2a}
\end{subfigure}%
\begin{subfigure}{.5\textwidth}
  \includegraphics[width=.8\linewidth]{Fig2b.eps}
  \caption{$h=11$ meV, $\lambda=18$ nm}
  \label{Fig2b}
  \end{subfigure}
 \vskip \baselineskip \vspace*{0.5cm}
   \begin{subfigure}{.5\textwidth}
  \includegraphics[width=.8\linewidth]{Fig2c.eps}
  \caption{$h=100$ meV, $\lambda=6$ nm}
  \label{Fig2c}
\end{subfigure}%
\begin{subfigure}{.5\textwidth}
  \includegraphics[width=.8\linewidth]{Fig2d.eps}
  \caption{$h=11$ meV, $\lambda=6$ nm}
  \label{Fig2d}
\end{subfigure}
\caption{ Superconducting critical temperature $T_c$ as a
function of the S layer thickness $d_s$ within three different magnetic configurations. The different panels (a), (b), (c), (d) correspond to different values of the exchange energy and spiral period. The large and small exchange field cases are characterized by the typical values $h=100$ meV and $h=11$ meV, respectively. The long and short spiral period situations correspond to $\lambda=18$ nm and $\lambda=6$ nm, respectively.\label{Fig2}}
\end{figure*}

We have presented a systematic study of the superconducting critical temperature $T_c$ of the S/M bilayer  with a comparison for the three configurations: a spiral parallel to the S/M interface ($\cos \alpha=0$), of a spiral inclined to the interface normal by the angle $\alpha=19.5^{\circ }$ (i.e., $\cos \alpha \approx 0.94$), and the case of   uniform magnetization . In this last case,  the superconducting properties are independent of the angle $\alpha$ and $Q=0$. These three configurations allow us to compare  two possible switching behaviors of the SSV,  both of which may be implemented experimentally.
If the S film is covered by a bulk or etched crystal of a spiral magnet of the B20 family, the change of $T_c$ may be achieved by the switch between parallel and inclined (maybe perpendicular) spiral. Alternatively, in the case of the proximity with a M film characterized by a strong in-plane
magnetic anisotropy (as in Ho, Er, or thin B20 family films), for which the
spiral vector is always perpendicular to the film plane, the $T_c$ switch may be
realized by the collapse to a uniform magnetization.

In Figs. \ref{Fig2}, the dependencies of $T_c$ on the superconductor thickness $d_s$ are first analyzed for different spiral configurations.
Fig. \ref{Fig2a} shows that for a strong exchange field,
the critical temperature $T_c$ for a parallel spiral is larger than that obtained for a uniform magnetization. This can be easily understood by the effective space averaging of the exchange field at the scale of the size of a Cooper pair. In contrast, for the inclined
spiral,  $T_c$  is (slightly) reduced in comparison to the uniform case because of the draining of the Cooper pairs from the S layer
into the open long-range triplet channel. This figure illustrates that the S/M proximity effect is driven by two competing mechanisms which affect the value of $T_c$: i) an effective spatial averaging of the exchange field (mostly governed by the wave vector $Q$ of the spiral), and ii) the appearance of the LRTC (controlled by the spiral angle and the exchange amplitude), which tends to suppress $T_c$.

In fact, the relative role of these two mechanisms depends on the relation between all characteristic wave vectors (including the exchange quasi-momentum $k_h$ defined from the exchange energy $h$, and the superconductor quasi-momenta related to the coherence lengths in the S and M layers), so that the resulting behavior for $T_c$ turns out to be a problem. depending on many parameters.  Our calculations reveal, as seen in Fig. \ref{Fig2b} , that the consideration of a smaller value for the exchange field $h$ than taken in Fig. \ref{Fig2a} can implement an opposite
order for the $T_c$ values, since the  $T_c$ for an inclined spiral is now larger than in the uniform case.
It can also be seen in Fig. \ref{Fig2b} that the spin-valve effect  for the set of parameters related to MnSi \cite{Bauer} may become giant,  reaching $\delta T_c \sim 1.6-1.8$ K depending on the switch to the inclined spiral or uniform magnetization.
An even larger spin-valve effect is displayed in Fig. \ref{Fig2d}, where both a low exchange field and a short spiral period are considered. This case can motivate a direction in the search for a suitable magnetic compound with the largest possible spin-valve effect. As expected, the strongest effects resulting from the S/F interplay may be found in the regime where all characteristic length scales (the S and F superconducting coherence lengths, the spiral wavelength $\lambda$, and the magnetic length $\xi_h=k_h^{-1}$) are of the same order.

To understand the behavior according to the two mechanisms i) and ii) we have highlighted, it is important to recall that the long-range and short-range triplet superconducting correlations are characterized by quite different dependencies on the exchange field, which are best discriminated at strong fields: the former has a $h$-independent coherence length $\sim (k_\omega^2+Q^2)^{-1/2}$, while the latter can be associated with the length scale $\sim (k_\omega^2+k_h^2)^{-1/2}$, which is roughly inversely proportional to $\sqrt{h}$.
The physical reason for this discrepancy is that the LRTC are not subjected to the depairing influence of the exchange field, so that both electrons in the Cooper pair can have the same spin projection to the field direction. Therefore, an increasing exchange field favors the draining of the Cooper pairs into the LRTC channel rather than into the short range triplet channel with $S_z=0$. At low fields, the LRTC appear as a correction in comparison with the predominantly generated short-range correlations, and thus hardly affect the superconducting properties of the M/S bilayer.
In the low-exchange field regime, the $T_c$ dependence on the spiral direction is mainly determined by the averaging mechanism:
when the direction of the penetration of the Cooper pairs into the magnet is different from  the direction of averaging, this averaging becomes more effective, and then $T_c$ proved to be suppressed to a lesser extent, cf. Fig. \ref{Fig2b}.

%%%%%%%%%%%%%%Fig.3

\begin{figure*}[htb]
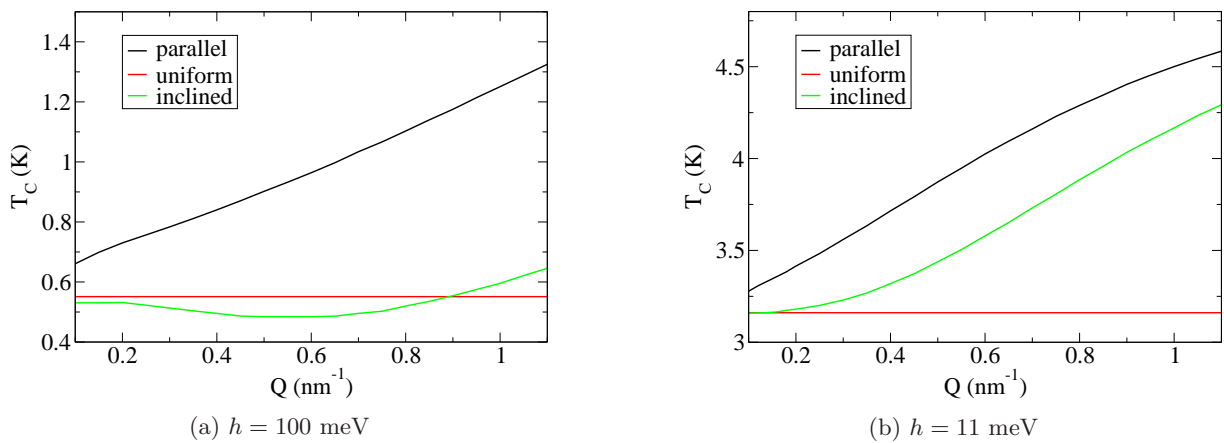

\begin{subfigure}{.5\textwidth}
  \includegraphics[width=.8\linewidth]{Fig3a.eps}
  \caption{$h=100$ meV}
  \label{Fig3a}
\end{subfigure}%
\begin{subfigure}{.5\textwidth}
  \includegraphics[width=.8\linewidth]{Fig3b.eps}
  \caption{$h=11$ meV}
  \label{Fig3b}
\end{subfigure}
\caption{
Superconducting critical temperature $T_c$ as a function of the spiral wave vector $Q$ within three different magnetic configurations, for (a) large and (b) small exchange energy $h$ (i.e., $h=100$ meV and $h=11$ meV, respectively). Here, we consider the layer thicknesses $d_f=40$ nm and $d_s=21$ nm. \vspace*{0.5cm}
}
\label{Fig3}
\end{figure*}

Interestingly, this averaging effect on $T_c$ may be even more pronounced in magnitude than that accompanied by the LRTC creation. It has already been noticed \cite{PugachPRB2009} that the scale of spatial modulation of superconducting properties in the junction plane may differ from that out of the plane. In the present study, one can also see the similarity between
the integrated energetic characteristics such as  $T_c$ and the Josephson current.
In the case of $T_c$,
the out-of-plane modulation scale is weakened due to the competition with the proximity effect leakage mechanism occurring in the same direction, so that out-of-plane modulations play quantitatively a lesser  role than the in-plane.
If one ignores the LRTC, one can understand this effect as follows: For a perpendicular or slightly inclined spiral, the Cooper pairs encounter, when penetrating into the M layer, an almost uniform exchange field and feel its modulations weakly as they propagate. In contrast, for a magnetic spiral parallel to the interface, they experience an exchange field averaged out over the full Cooper pair size extension.

As already remarked, the low exchange field value in the compound MnSi  yields, {\it a priori}, an important spin-valve effect.
As is clear from the previous discussion, another way to enhance the switching effect would be to prefer magnetic materials with strong spatial modulations, as the mean exchange averaging becomes more effective for a short spiral period. This situation is encountered,  for example, in Ho with its relatively short spiral period $\lambda=6$ nm. The corresponding $T_c$ dependencies for the different spiral configurations are respectively displayed in Figs. \ref{Fig2c} and \ref{Fig2d}  for relatively large (100 meV) and small (11 meV) exchange energies \cite{remark}.
%(we have chosen  here $h=100$ meV).
%because we believe \cite{remark} that pure 4f metals still have strong exchange.
From the comparison between Figs. \ref{Fig2a} and \ref{Fig2c} (large exchange field case), or between \ref{Fig2b} and \ref{Fig2d} (small exchange field case), it is clear that the averaging mechanism leads to important quantitative changes in the $T_c$ dependencies. Moreover, this effect turns out to be magnified when the magnetic length $\xi_h$ becomes closer to the superconducting length $\xi_f$ (a situation encountered in Figs. \ref{Fig2b} and \ref{Fig2d}).

From now on, we will consider a S/M bilayer with a fixed S thickness ($d_s=21$ nm; this value is chosen as more or less optimal to exhibit the dependencies) and analyze the $T_c$ variations with respect to various other model parameters. We show in Fig. \ref{Fig3} that
the interplay between the averaging and the LRTC mechanisms is also exhibited in the dependence of the superconducting critical $T_c$ on the spiral wave vector $Q=2 \pi/\lambda$. For a large exchange field, we clearly see in Fig. \ref{Fig3a} an extremum in the $T_c$ dependence of the inclined spiral. It is the result of the competition between the  $T_c$ suppression by the LRTC, which is more
important at a long spiral, i.e., at small $Q$, and the effective averaging mechanism, which leads to the growth of $T_c$ at a short spiral period, i.e., at large $Q$. Note that this competition does not occur in the case of a spiral parallel to the interface, for which the LRTC is absent: $T_c$ displays a monotonic increase as a function of $Q$ due to the more efficient averaging effect when reducing the characteristic spatial modulation of the exchange field.
In contrast, at a smaller exchange field, the LRTC creation mechanism becomes less relevant, so that the extremum for the inclined configuration shifts to much smaller values of $Q$ as revealed in Fig. \ref{Fig3b}. In this situation the averaging effect dominates
for most spiral periods.

\begin{figure*}[htb]
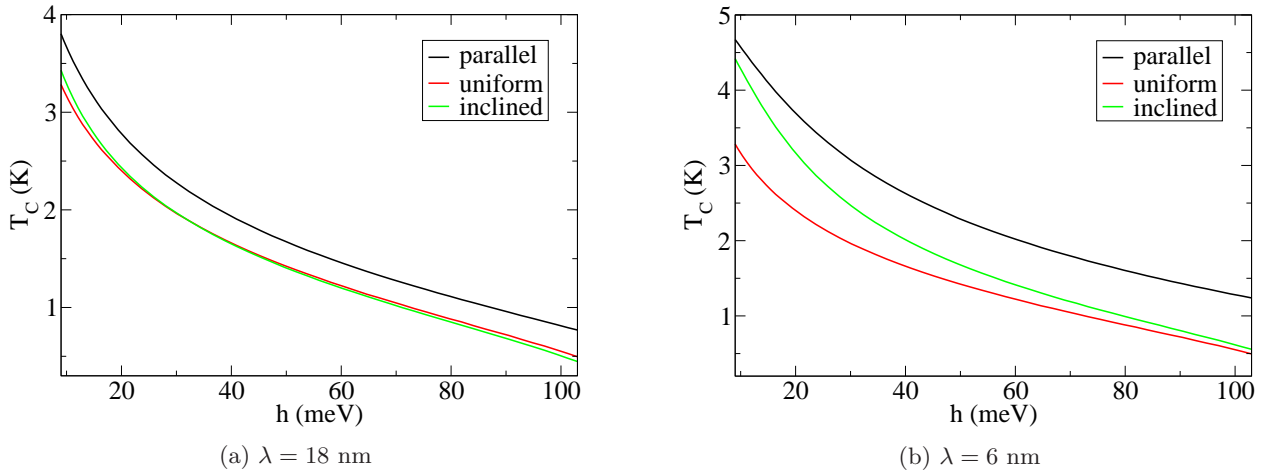

\begin{subfigure}{.5\textwidth}
  \includegraphics[width=.85\linewidth]{Fig4a.eps}
  \caption{$\lambda=18$ nm}
  \label{Fig4a}
\end{subfigure}%
\begin{subfigure}{.5\textwidth}
  \includegraphics[width=.85\linewidth]{Fig4b.eps}
  \caption{$\lambda=6$ nm}
  \label{Fig4b}
\end{subfigure}
\caption{
Superconducting critical temperature $T_c$ as a
function of the exchange energy $h$ within three different magnetic configurations, (a) for large spiral spatial period $\lambda=18$ nm (as in MnSi), and (b) for short spiral spatial period $\lambda=6$ nm (as in Ho). We have taken $d_s=21$ nm and a large M layer thickness $%
d_f=40$ nm$ \gg\protect\xi_f$.
}
\label{Fig4}
\end{figure*}

To make clear the role of the exchange energy in the interplay of the two
mechanisms governing the behavior of $T_c$ that we have emphasised, we present in Fig. \ref{Fig4a} the dependence of $T_c$ on $h$ for a long spiral period (case of MnSi with $\lambda=18$ nm). It is seen that the
$T_c$ in the different configurations
display a similar (roughly parallel) decreasing tendency. In the low exchange field regime, the inclined case provides a slightly higher $T_c$ than in the uniform case, whereas at higher exchange values it yields a lower $T_c$.
 In the inclined spiral configuration, the growth of $h$ entails an increasing role of the LRTC in the proximity effect, which efficiently weaken the superconductivity of the S layer.
At a smaller spiral period, as in Ho ($\lambda=6$ nm), the differences in the behavior of $T_c(h)$  between the uniform and inclined configurations is more pronounced,  see Fig. \ref{Fig4b}. The lowest $T_c$ is then always obtained for the uniform case.

\begin{figure*}[htb]
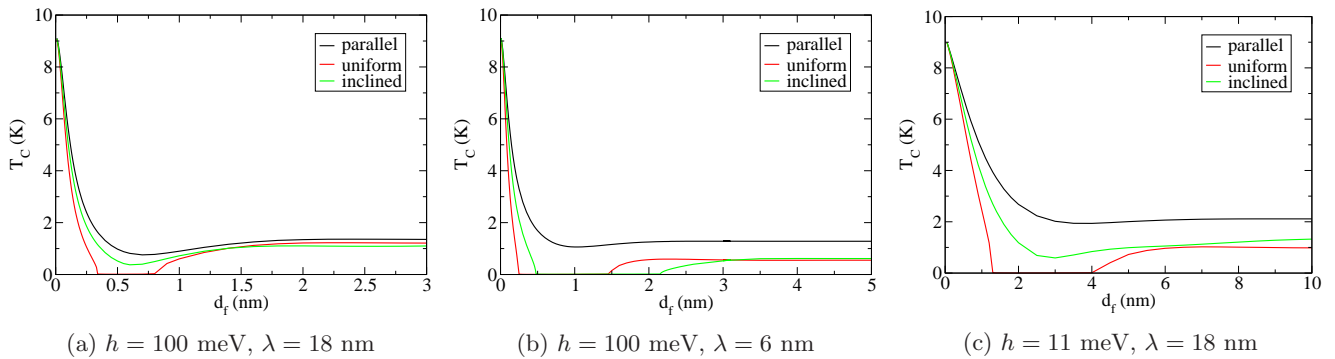

\begin{subfigure}{.33\textwidth}
%\hspace*{-1cm}
  \includegraphics[width=0.95\linewidth]{Fig5a.eps}
  \caption{$h=100$ meV, $\lambda=18$ nm}
  \label{Fig5a}
\end{subfigure}%
%\hspace*{-1cm}
\begin{subfigure}{.33\textwidth}
  \includegraphics[width=0.95\linewidth]{Fig5b.eps}
  \caption{$h=100$ meV, $\lambda=6$ nm }
  \label{Fig5b}
\end{subfigure}%
\begin{subfigure}{.33\textwidth}
  \includegraphics[width=0.95\linewidth]{Fig5c.eps}
  \caption{$h=11$ meV, $\lambda=18$ nm}
  \label{Fig5c}
\end{subfigure}%
\caption{
Superconducting critical temperature $T_c$ as a
function of the M layer thickness $d_f$ within three different magnetic configurations, (a) for large exchange energy $h=100$ meV,
large spiral spatial period $\lambda=18$ nm and $d_s=21.4$ nm, (b) large exchange energy $h=100$ meV, short
spiral spatial period $\lambda=6$ nm and $d_s=$ 21 nm, and (c) for small exchange energy $h=11$ meV, large spiral spatial period $\lambda=18$ nm and $d_s=18.5$ nm (MnSi case).
}
\label{Fig5}
\end{figure*}

The finite thickness of the M layer taken into account in our model calculations allows us to consider the phenomenon of re-entrant superconductivity according to the spiral configuration, which is studied in some detail in Figs. \ref{Fig5}. We point out that if one magnetic configuration provides re-entrant behavior, i.e., $T_c=0$ in
some range of thickness $d_f$, a full switch of superconductivity may then be achieved in this range with the change of the magnetic configuration.
With the decrease of
the magnetic exchange energy (compare Fig. \ref{Fig5a} and Figs. \ref{Fig5b}, \ref{Fig5c}) the period of the  $T_c$
oscillation becomes larger, as one could expect. It thus creates better conditions
for the experimental implementation of such re-entrant behavior, by providing wider regions
of $d_f$ where $T_c=0$ for a given magnetic configuration and nonzero $T_c$ (in the Kelvin range) for another magnetic configuration. Note that the nature of the magnetic configuration yielding a vanishing  $T_c$ in a given $d_f$ range also depends on the magnetic exchange amplitude, in accordance with Figs. \ref{Fig4}.
 For thin films grown with a strong in-plane anisotropy a possible switch may be realized by commuting the magnetic system from a spiral
perpendicular to the layer to a uniform magnetization configuration. This  may be achieved for a thickness smaller than the spiral period under  application of a relatively small magnetic field. From this analysis, one may
conclude that the compounds of the B20 family seem preferable for the observation of
controlled re-entrant superconductivity phenomenon and for the realization of the full switch of $T_c$ to zero.
%%%%%%%%%%%%%%%%%%%%%%%%%%%%%%%%%%%%%%%%%%%%%%%%%%%%%

\begin{figure*}[htb]
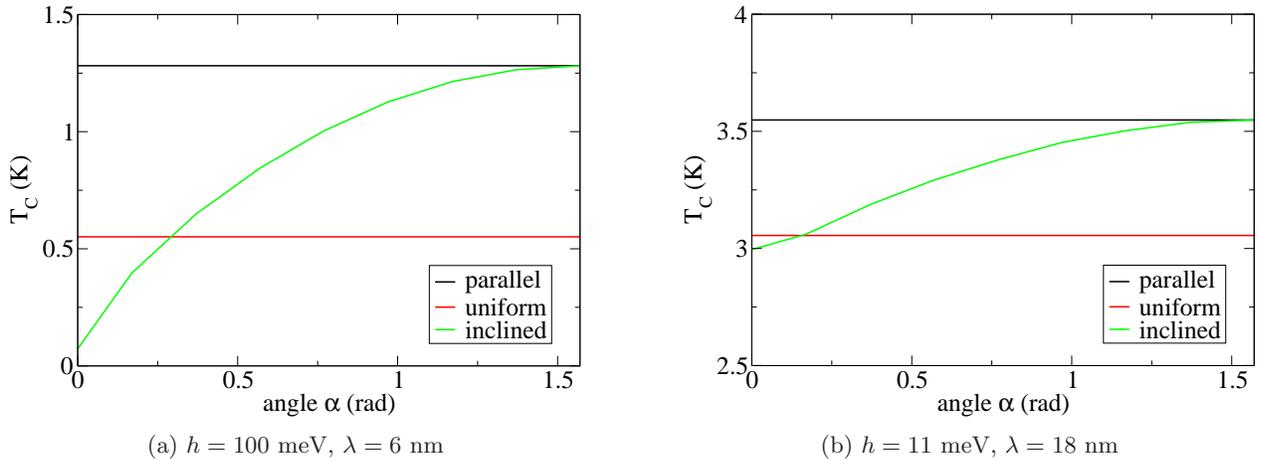

\begin{subfigure}{.5\textwidth}
  \includegraphics[width=.85\linewidth]{Fig6a.eps}
  \caption{$h=100$ meV, $\lambda=6$ nm}
  \label{Fig6a}
\end{subfigure}%
\begin{subfigure}{.5\textwidth}
  \includegraphics[width=.85\linewidth]{Fig6b.eps}
  \caption{$h=11$ meV, $\lambda=18$ nm}
  \label{Fig6b}
\end{subfigure}
\caption{
Superconducting critical temperature $T_c$ as a
function of the angle $\alpha$ between the spiral vector $Q$ and the S layer normal
within three different magnetic configurations, (a) for large exchange $h=100$ meV and short spiral spatial period $\lambda=6$ nm, and (b) for small exchange $h=11$ meV and long spiral spatial period $\lambda=18$ nm (as in MnSi). We have taken equal layers thicknesses $d_s=d_f=21$ nm. The curves in the uniform and parallel cases are kept here as reference lines.
}
\label{Fig6}
\end{figure*}

Finally, our model calculations also allowed us to study the $T_c$ dependence on the angle $\alpha$ between the spiral vector ${\bf Q}$ and the S/M interface normal, see Figs. \ref{Fig6} with two different sets of parameters, one suitable for MnSi (Fig. \ref{Fig6b}) and the other one for Ho (Fig. \ref{Fig6a}). In the two cases, the same general trend is found, namely a monotonic growth of $T_c$ with the angle $\alpha$, with an initial $T_c$ lower than the $T_c$ obtained in the uniform magnetization case. This stronger suppression of superconductivity for a spiral perpendicular to the film is due to the presence of the LRTC, whose generation is controlled  by the angle (with a maximal production for $\alpha=0$). The comparison between Fig. \ref{Fig6a} and Fig. \ref{Fig6b} shows that the LRTC plays a role at lower exchange field and at larger spiral period only within a small range of angles $\alpha$ for  which we encounter the unusual situation where an applied magnetic field magnetizing the film uniformly may lead to an increase of the $T_c$ of the bilayer.

\section{Conclusion}

\label{Conclusion}

To summarize, we have studied in detail a superconducting spin valve consisting of a superconducting film covered by a spiral magnet characterized by multiple equilibrium configurations. We have considered spiral magnets of different types: Ho and Er, or B20 family compounds, both crystals and films. These SSVs may be controlled by
biased or pulsed external magnetic field, producing magnetization reversal or magnonic relaxation resulting in
reorientation of the  magnetic spiral. They present not only a fundamental theoretical interest, but also a technological one, as exemplified by the earlier experimental works \cite{DiBernardo2015,Satchell2017} on Er or Ho bilayers and recent experimental efforts to create MnSi based SSVs \cite{Grefe}.

In comparison with the theoretical work developed previously in Ref. \onlinecite{Pugach2017,VolkovAnishchanka} on the same model, we have presented here an
extended derivation including a number of significant improvements: i) an arbitrary
value of the exchange energy, suitable for MnSi for example, is considered, whereas a large  value was assumed previously \cite{Pugach2017}; ii) the magnetic spiral vector can be inclined at an arbitrary angle with respect to the interface (for simplicity, only parallel and perpendicular spirals were considered in Ref. \onlinecite{Pugach2017}); iii) the finite thickness of the magnetic layer is taken into account, in contrast to the previous work \cite{Pugach2017} that assumed a semi-infinite M layer.

Besides providing a more realistic description of the possible experimental conditions and justifying some simplifications made in the previous work, this enlargement of the parameter space reveals quite different qualitative behaviors for the spin valve depending on the spiral configuration. In particular, we have found that the superconducting critical temperature may significantly vary as a function of the spiral direction also at moderate or low exchange energies. In this parameter regime, the superconducting correlations resulting from the proximity effect between the superconducting film and the chiral magnet are primarily short-range, so that the sensitivity of $T_c$ is only due to
the averaging of the exchange field on the scale of the Cooper pairs. It turns out that this averaging effect is much more efficient when the direction, perpendicular to the interface, of the penetration of the Cooper pairs  does not coincide with the direction of the magnetic spatial inhomogeneity. As a result, the parallel spiral configuration always yields the highest $T_c$, as confirmed by our numerical calculations.

In contrast, for large exchange energies a different mechanism for the sensitivity of $T_c$ is at play: for a spiral direction orthogonal to the interface the production of long-range superconducting triplet correlations opens one more channel for the leakage of the Cooper pairs from the superconducting film, thus leading to an efficient decrease of $T_c$.
Therefore, the behavior of the studied superconducting spin valve is driven by two different antagonistic mechanisms. Our quantitative analysis of the reentrant superconductivity phenomenon in the bilayer suggests the switching behavior of the superconducting spin valve should be optimized  at moderately low exchange energies, where the averaging mechanism comes to dominate that of  long-range triplet correlations.

The most interesting result, which should stimulate further experimental investigation, optimization, and manufacturing of such structures, is that  the spin valve effect  may yield a $\delta T_{c}$ within the Kelvin range in MnSi-Nb bilayers, i.e., may be giant. Therefore, it seems generally preferable to manufacture SSVs with MnSi crystal-Nb bilayers. The intrinsic solution of the half-select problem and the possibility of magnonic control, that is  non-destructive for the superconductivity, make such spin valves appealing for applications in cryogenic nanoelectronics.

\subsection{Acknowledgements}

N.P. thanks D. Menzel, V. Dyadkin, and A. Bauer for helpful discussions about the
B20 family properties. The French-Russian collaboration has benefited from the Visiting Scientist
Program of the Centre de Physique Th\'{e}orique de Grenoble-Alpes, and from the invited researcher program of CNRS.
The analysis of B20 family based SSVs was financially supported by the Ministry of Science and Higher Education of the Russian Federation, Megagrant project N 075-15-2022-1108.
 N.P. thanks also the Project Mirror Labs of the HSE University for the support of the analysis of 4f metals - superconductor bilayers.

\section*{Appendix}

 \label{Appendix}
 In this Appendix, we present a detailed derivation of the key-quantity $W$ [see Eq. \eqref{W}], which determines the dependencies of the superconducting critical temperature on the different model parameters.

For an inclined spiral, the magnetic structure is nonuniform both in $OZ$ and $OX$ directions, so that the different superconducting correlations components in the linearized Usadel Eqs. \eqref{UU0} are expected to be in general functions of both $x$ and $z$ spatial variables, and the
 Laplacian operator reads $
\nabla ^{2}=\frac{\partial ^{2}}{\partial x^{2}}+\frac{\partial ^{2}}{%
\partial z^{2}}$.

 In the orthogonal case ($\alpha =0$), the
spiral vector $\mathbf{Q}$ lies in the $OZ$ direction, and the
magnetic field direction is described by the rotation matrix $M_{z}$ around $%
OZ$ axis with the rotation phase $\beta =Qz$: $\mathbf{{h}=}h(\cos \beta
,\sin \beta ,0)^{T}=M_{z}\left( h,0,0\right) ^{T}\mathbf{.}$ The linearized
Usadel equations take the form in this case:
\begin{eqnarray}
\left( D\nabla ^{2}-2\left\vert \omega \right\vert \right) f_{s} &=&2i\
sgn(\omega )h\left[ f_{x}\cos (Qz)+f_{y}\sin (Qz)\right] ,  \notag \\
\left( D\nabla ^{2}-2\left\vert \omega \right\vert \right) f_{x} &=&2i\
sgn(\omega )hf_{s}\cos (Qz),  \label{UU:1} \\
\left( D\nabla ^{2}-2\left\vert \omega \right\vert \right) f_{y} &=&2i\
sgn(\omega )hf_{s}\sin ( Qz).  \notag
\end{eqnarray}

From now on, we assume that the axis of the spiral is inclined in the $XOZ$ plane. This inclined
configuration may be obtained from the orthogonal case with the help of
the rotation matrix around the $OY$ axis by the angle $\alpha $ ($OZ$ axis
rotates towards the $OX$ axis by the angle $\alpha $)
\begin{equation}
M_{z}=\left(
\begin{array}{ccc}
\cos \beta & -\sin \beta & 0 \\
\sin \beta & \cos \beta & 0 \\
0 & 0 & 1
\end{array}
\right) , \,  M_{y}=\left(
\begin{array}{ccc}
\cos \alpha & 0 & -\sin \alpha \\
0 & 1 & 0 \\
\sin \alpha & 0 & \cos \alpha
\end{array}
\right)  \label{MzMy}
\end{equation}
where the inclined spiral vector is given by $\mathbf{{Q}=}M_{y} (0,0,Q)^{T}$.
The phase of the exchange field vector rotation around the spiral vector
is then $\beta =\beta (x,z)=Q\left( -x\sin \alpha +z\cos \alpha \right)$. The
exchange field vector reads $\mathbf{{h}=}M_{y}M_{z}\left( h,0,0\right)
^{T}=h(\cos \alpha \cos \beta ,~\sin \beta ,~\sin \alpha \cos \beta )^{T}%
\mathbf{.}$

The unitary transformation of the triplet components vector $\mathbf{f}_{t}=(f_{x},~f_{y},~f_{z})^{T}$ to the vector $\mathbf{f}_{pm}=(f_{+},~f_{-},~f_{z})^{T}$ in the orthogonal case ($\alpha =0$) is given by $\mathbf{f}_{pm}=N\mathbf{f}_{t}$ with
the matrix $N$
\begin{eqnarray}
N=\left(
\begin{array}{ccc}
-\exp (i\beta ) & i\exp (i\beta ) & 0 \\
\exp (-i\beta ) & i\exp (-i\beta ) & 0 \\
0 & 0 & 1
\end{array}
\right). \label{fpm}
\end{eqnarray}
In the inclined case ($\alpha \neq 0$) all components of $\mathbf{f}_{t}$
 are nonzero in general. We may then introduce the vector $\mathbf
 {f}_{pm }=(f_{+},~f_{-},~f_{zz})^{T}$ given by the linear transformation $\mathbf{
f}_{pm }=NM_{y}^{-1} \mathbf{f}_{t}$, which transforms the linearized Usadel
Eqs. \eqref{UU:1} to their linear combination. The equation for the third triplet component $f_{zz}$ separates after the
transformation and we may choose $f_{zz}=0,$ that yields the additional
condition $f_{z}=f_{x}\tan \alpha$.

The explicit dependence on the $x$ variable has completely disappeared in the equations obeyed by the correlation functions $f_{s,\pm }(x,z)$. It can be shown that the latter functions are in fact independent of $x$. For this purpose, we follow the method of Ref. \onlinecite{Champel2005} by expanding the correlations as Fourier series, assuming that the system is periodic in the $OX$ direction: $f_{s,\pm }(x,z)\rightarrow
f_{s,\pm }(m,z)$ where $m$ is an integer number.  Since the equations are linear, the different
Fourier components $m$ are fully decoupled. The most energetically favorable
solution corresponds \cite{Champel2005} to the uniform one with $m=0$. The other solutions with $m=\pm 1,\pm
2...$ describe the cases of a superconducting order parameter which is spatially inhomogeneous even deeply in the S region
and are not realized physically. Finally, we find that the functions $f_{s,\pm }(x,z)$
obey the Eqs. (\ref{UU:2}) in the ferromagnet, with a form similar to the
orthogonal case, but  with the spiral vector $Q$ replaced by $Q\cos \alpha $ in the linear term.

The eigenvectors of the system (\ref{UU:2}) may be chosen for the eigenvalues $%
k_i=k_{0,\pm }$ as
\begin{equation}
\left(
\begin{array}{c}
u_{s}(k_i) \\
u_{-}(k_i) \\
u_{+}(k_i)
\end{array}
\right) =\left(
\begin{array}{c}
i\frac{k_i^{2}-k_{\omega }^{2}+2iQk_i\cos \alpha -Q^{2}}{2k_{h}^{2}} \\
-\frac{k_i^{2}-k_{\omega }^{2}+2iQk_i\cos \alpha -Q^{2}}{k_i^{2}-k_{\omega
}^{2}-2iQk_i\cos \alpha -Q^{2}} \\
1
\end{array}
\right). \label{Eigenvec:u}
\end{equation}
For a finite F layer we also have to take into account the wave reflected
from the free M layer interface solutions $\sim\exp (k_{i}z)$. The
eigenvectors $v_{j}$ representing this reflected wave may be obtained with the change $k_{i}\rightarrow -k_{i}$, considering that only the relation between the eigenvector components has a meaning
\begin{equation}
\left(
\begin{array}{c}
v_{s}(k_i) \\
v_{-}(k_i) \\
v_{+}(k_i)
\end{array}
\right) =\left(
\begin{array}{c}
i\frac{k_i^{2}-k_{\omega }^{2}+2iQk_i\cos \alpha -Q^{2}}{2k_{h}^{2}} \\
-1 \\
\frac{k_i^{2}-k_{\omega }^{2}+2iQk_i\cos \alpha -Q^{2}}{k_i^{2}-k_{\omega
}^{2}-2iQk_i\cos \alpha -Q^{2}}
\end{array}
\right) =\left(
\begin{array}{c}
u_{s}(k_i) \\
-u_{+}(k_i) \\
-u_{-}(k_i)
\end{array}
\right).   \label{Eigenvec:v}
\end{equation}
 Note that for $\cos \alpha =0$ (spiral parallel to the SM interface) the two functions $f_{\pm}$ are identical, thus indicating that only one type of triplet component (of short-range nature) is in fact produced \cite{Champel2005} in the hybrid structure in this special configuration.
In the inclined case, the system resolution necessarily involves the two different components $f_{+}$ and $f_{-}$ hinting at the coexistence of two types of triplet correlations in the structure.
Owing to the chirality, the short-range  components (depending on the exchange field $h$) of the
eigenvectors for the incident $\mathbf{u}$ and reflected $\mathbf{v}$ waves
display a
complicated symmetry $v_{\pm}=-u_{\mp}$,
including the chirality change ("+" to "-", and opposite). The reflected wave moves along the spiral backward and feels the opposite direction (clockwise or anticlockwise) of the magnetization rotation along the movement direction.
 It is worth stressing that the combination of chirality and finite thickness for the magnet leads to a computational complexity related to the doubling of the coefficient amplitudes needed, which was absent in  previous publications \cite{Champel2005,VolkovAnishchanka,Halasz2009,Halasz2011,Wu2012,USA2014,Halterman2016,Pugach2017}.

The general form of the solution for the finite F layer is written down in Eq. \eqref
{GenSol}. Let us denote $R_{i}\equiv A_{i}-B_{i}$ and  $S_{i}\equiv
A_{i}+B_{i}$. From the boundary conditions (\ref{BCfree}) at $z=d_{f}$, it follows the equations
\begin{eqnarray}
\sum\limits_{i}\left[ R_{i}\cosh (k_{i}d_{f})-S_{i}\sinh (k_{i}d_{f})\right]
U_{s,- }(k_{i}) &=& 0,  \nonumber \\
\sum\limits_{i}\left[ S_{i}\cosh (k_{i}d_{f})-R_{i}\sinh (k_{i}d_{f})\right]
U_{+}(k_{i}) &=& 0, \hspace*{0.5cm}
\label{BCdf2}
\end{eqnarray}
where we have introduced $U_{s}(k_{i}) = u_{s}(k_{i})k_{i}$ and
\begin{eqnarray}
U_{\pm}(k_{i}) &= &V_{\pm}(k_{i})k_{i}+iQ\cos \alpha V_{\mp}(k_{i}).
\\
\quad
V_{\pm}(k_{i}) &=&
u_{+}(k_{i}) \pm u_{-}(k_{i}), \nonumber
\label{defUV}
\end{eqnarray}

In the S layer the spatial non-uniformity is only provided by the
nonuniform magnetization existing along the SM interface. It thus seems to be natural to
use $\beta =\beta (x,0)$ inside the S region.  So the transformation to apply on the different superconducting components has the form
$\mathbf{f}_{pm }^S=N(\beta (x,0))M_{y}^{-1}
\mathbf{f}_{t}^S$, that yields the Usadel equations in the S region
\begin{eqnarray}
\left( \nabla ^{2}-\frac{2\omega }{D_{s}}\right) f_{s}^{S} =-2\pi \ \frac{%
\Delta }{D_{s}},  \notag \hspace*{3.3cm} \\
\left[ \nabla ^{2}\pm 2iQ\sin \alpha \frac{\partial }{\partial x}-Q^{2}\sin
^{2}\alpha -\frac{2\omega }{D_{s}}\right] f_{\pm }^{S} =0.  \hspace*{0.5cm}\label{UUs1}
\end{eqnarray}
As explained above, the $x$ dependence has been dropped out in the transformed components $f_{s,\pm}$,
so that we get the equations in the S region
\begin{eqnarray}
\left( \frac{\partial ^{2}}{\partial z^{2}}-\frac{2\omega }{D_{s}}\right)
f_{s}^{S} =-2\pi \ \frac{\Delta }{D_{s}},  \notag \hspace*{3cm} \\
\left[ \frac{\partial ^{2}}{\partial z^{2}}-Q^{2}\sin ^{2}\alpha -\frac{%
2\omega }{D_{s}}\right] f_{\pm }^{S} = 0,  \hspace*{2.5cm}\label{UUs2}
\end{eqnarray}
and  $f_{zz}^{S}=0$, as it was also chosen in the ferromagnet. We may search the solutions
for the triplet components $f_{\pm }^{S}$ satisfying the boundary condition (\ref{BCfree}) at
the free interface $z=-d_s$  in the form:
\begin{equation}
f_{\pm }^{S}=C_{\pm }\frac{\cosh [k_{s}(z+d_s)]}{\sinh (k_{s}d_s)},  \label{SolS}
\end{equation}
with the wave vector $k_{s}=\sqrt{Q^{2}\sin ^{2}\alpha+ 2\omega /D_{s}}$. The boundary conditions (\ref{BC:SM}) at the SM
interface ($z=0$)  read
\begin{eqnarray}
\sum\limits_{i}R_{i}V_{+}(k_{i})+S_{i}U_{+}(k_{i})\xi _{f}\gamma _{c} &=&0,
\notag \\
\sum\limits_{i}S_{i}V_{-}(k_{i})+R_{i}U_{-}(k_{i})\xi _{f}\gamma _{c} &=&0,
\label{RSeqs}
\end{eqnarray}
where $\gamma _{c}= \gamma _{b}+\gamma \coth (k_{s}d_{s})/\xi _{s}k_{s}$.
For the singlet component, we finally get the two equations:
\begin{eqnarray}
\frac{\partial }{\partial z}f_{s}^{S}(0) &=&-\gamma \frac{\xi _{f}}{\xi _{s}}%
\sum\limits_{i}R_{i} k_i u_{s}(k_{i}), \nonumber \\
f_{s}^{S}(0) &=&\sum\limits_{i}S_{i}u_{s}(k_{i})+\gamma
_{b}R_{i}k _i u_{s}(k_{i})\xi _{f}. \label{final}
\end{eqnarray}
The latter equations can then be recast in the form of the reduced boundary
condition (\ref{BCW}) with the expression (\ref{W})
for the value of the quantity $W$. The coefficients $R_{i},S_{i}$ are determined by
using Eqs. (\ref{BCdf2}) and (\ref{RSeqs}).

\end{document}